# The plane stress state of residually stressed bodies: a stress function approach

Michael Moshe* and Eran Sharon†
*Racah Institute of Physics, The Hebrew University, JerusaleM*

Raz Kupferman
*Einstein Institute of Mathematics, The Hebrew University, Jerusalem*

The stressed state of flattened thin elastic sheet, as well as that of translationally symmetric 3D solids - are effectively 2D problems. This paper study equilibrium state-of-stress in metrically-incompatible 2D elastic materials. The solution is represented by a scalar stress function, generalizing the Airy stress function, which is determined by geometric compatibility conditions. We develop a perturbative approximation method for solving this stress function, valid for any constitutive relation. We apply the method for the case of a Hookean solid to solve prototypical examples in which the classical Airy approach is either inaccurate or inapplicable. Results are shown to agree well with numerical results obtained in previous works.

## I. INTRODUCTION

A classical problem of practical importance in elasticity is finding the equilibrium state of an elastic body. Stressed states usually arise as a response to external forcing. Another class of settings in which stresses are present at equilibrium is when the intrinsic geometry of the material is incompatible with the ambient Euclidean space. This incompatible geometric structure is usually induced by inelastic deformations. Stresses present in the absence of external constraints are called residual stresses. Residual stress is common, for example, in systems subject to thermal gradients, in bodies with defects [1], and in natural tissues that undergo non-uniform growth [2].

It was recently shown that defective materials [3], as well as growing natural tissues [4], can be modeled within a geometric formulation of elasticity. In this formulation the local rest distances between material elements are described by a reference metric tensor $\bar{\mathfrak{g}}$. This geometric formulation of elasticity, which is suitable for the description of large inealstic deformations, is also relevant for the description of Non-Euclidean Plates (NEP). NEP is a thin elastic sheet that is uniform along its thickness, and described using a 2D reference metric that is not necessarily euclidean. The stressed state of 3D materials that are symmetric under translations along an axis can be viewed as the state of a very thick NEP, or alternatively, as the state of a flattened thin NEP. Such symmetric materials can be effectively described as a NEP with a 2D reference metric.

The stressed state of defective materials, as well as that of pre-buckled NEP of finite thickness, are of practical importance in the study of material's properties. The geometric formulation of elasticity has for advantage that it is formulated using entities of physical and geometric significance (e.g., curvatures and parallel transport). A disadvantage of this approach is its strong nonlinearity, which makes it hard to apply to practical applications. As an example, the exact elastic equilibrium equation of thick/flattened NEP was derived [5], but no general analytical methods were developed for solving it.

Common approaches to residually-stressed bodies, which are limited to small inelastic deformations (or in mathematical language, to weak geometric incompatibility) are not suitable to many of the cases of interest of NEPs. For example, growing natural tissues and active materials (e.g., [6, 7]) are two classes of systems that may involve large inelastic deformations. In order to elucidate the complex patterns generated by those systems one has to rely on models that go beyond a weak incompatibility regime.

In this work we develop an approximation method for solving the equilibrium plane-stress state of non-Euclidean plates, obtained in [8]. We show that the solution for the stress can be represented in terms of a scalar function, the Incompatible Stress Function (ISF), which generalizes the Airy stress function used in linear elasticity [1]. The representation of the stress by the ISF is an analytical property that does not rely on any approximation, and capture both nonlinearity and incompatibility. Under a specified constitutive law, one obtains a representation for the actual metric at equilibrium in terms of the ISF. This procedure is valid for any constitutive law. Geometric compatibility conditions satisfied by the actual metric, along with boundary conditions, determine the ISF.

We consider examples that include both simply-connected and multiply-connected domains. In the latter case, the geometric compatibility conditions provide in a natural way additional constraints, which have no immediate counterpart in the classical Airy stress function formulation.

To find the ISF we use a perturbative approach in which the small (formal) parameter is a measure of geometric incompatibility. To lowest order one obtained a linear fourth-order equation for the ISF, which can be viewed as a geometric generalization of the biharmonic equation satisfied by the Airy stress function. Higher-order corrections can be obtained systematically; in this paper we demonstrate how to derive second-order corrections.

* michael.moshe@mail.huji.ac.il
† erans@mail.huji.ac.il

We apply our method on two prototypical examples. The first example models a thick disc with a single disclination line parallel to the $z$ axis - a problem that could also be solved using the Airy approach. The reference metric of this geometry prescribes a delta-function singularity of reference Gaussian curvature. The thin plate limit of such discs was studied in [9]. The second example models a material with uniformly constant reference Gaussian curvature - positive or negative. The plane stress states of both examples were presented and solved numerically in [10] in the context of non-Euclidean thin plates embedded in the Euclidean plane. According to this interpretation, our first example models a flattened cone, while the second one models a flattened sphere (or a flattened surface of constant negative Gaussian curvature) Fig. 1

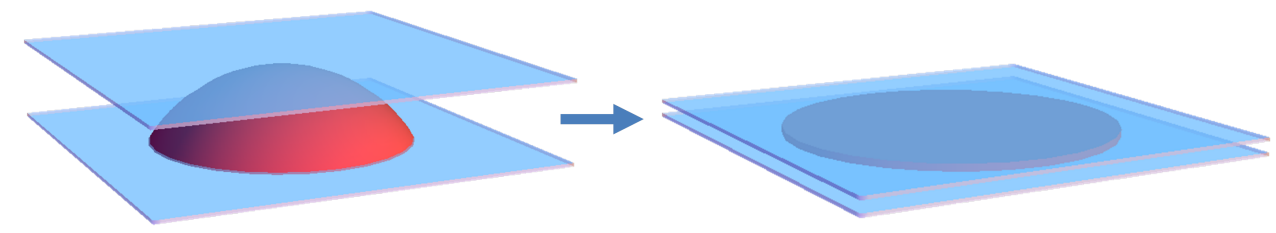

FIG. 1: An illustration of the plane stress state of a NEP, having a reference metric of constant positive Gaussian curvature.

Our results are compared to the numerical solutions of the fully nonlinear problem presented in [10]. For the problem of a single disclination line we also compare our results to that obtained by the classical Airy stress approach.

Finally, in addition to its relevance for the mechanics of NEP (such as for calculation of their buckling threshold ), the formalism is relevant for other cases of 3D axially symmetric systems, i.e., systems whose state-of-stress is essentially two-dimensional. These include problems of rods that are residually stressed due to growth [], or due to thermal gradients.

## II. INCOMPATIBLE ELASTICITY

In certain geometric formulation of elasticity theory, an elastic body is modeled as a Riemannian manifold $\mathcal{B}$ equipped with a reference metric $\bar{\mathfrak{g}}$ [5], which describes local rest distances between adjacent material elements. An elastic body is commonly assumed stress-free in the absence of external constraints. This statement is equivalent to saying that the reference metric $\bar{\mathfrak{g}}$ is Euclidean. In many cases of interest, however, the reference metric is non-Euclidean, leading to a theory of incompatible elasticity. Incompatible elasticity was developed in the 1950s in the context of crystalline defects; it has attracted renewed interest in recent years in other contexts, such as thermo-elasticity [11], growth dynamics [7, 12], differential swelling [4, 6, 13–16], and macro-molecules self assembly [17]. It should be noted that in general, material manifolds may be endowed with properties other than just a metric, which is a particular case of a section of a fiber bundle [18]. The present work assumes a homogeneous and isotropic medium, fully described by its metric properties.

A configuration of an elastic body is an embedding of $\mathcal{B}$ in the ambient Euclidean space (the space manifold $\mathcal{S}$). Every configuration induces on $\mathcal{B}$ a metric, $\mathfrak{g}$, which quantifies actual distances between adjacent material elements ($\mathfrak{g}$ is the pullback of the Euclidean metric). The elastic strain tensor is the discrepancy between the actual metric and the reference metric,

$$u = \frac{1}{2}\left(\mathfrak{g} - \bar{\mathfrak{g}}\right). \tag{1}$$

Note that this definition of the strain tensor is purely geometric and involves no linearization.

The elasto-static model is fully determined by a constitutive law, or in the case of a hyper-elastic material, by an energy functional. This energy functional is an additive measure of local strains. In first-grade elasticity, the energy density is assumed to only depend on the first derivative of the configuration. Assuming frame indifference, the energy functional can be written in terms of the actual metric,

$$E = \int W(\mathfrak{g}; \bar{\mathfrak{g}})\, d\mathrm{Vol}_{\bar{\mathfrak{g}}} \ , \tag{2}$$

where $d\mathrm{Vol}_{\bar{\mathfrak{g}}}$ is the Riemannian volume element, and $W$ is a non-negative energy density (viewed here as a function of the section of metric tensors) that vanishes at $x$ if and only if $\mathfrak{g}(x) = \bar{\mathfrak{g}}(x)$. Incompatibility manifests in that $\mathfrak{g}$ cannot be equal to $\bar{\mathfrak{g}}$ everywhere simultaneously.

It can be shown (in a way similar to [19]) that the configuration that minimizes the energy satisfies the equilibrium equations,

$$\bar{\nabla}_\mu \sigma^{\mu\nu} + \left(\Gamma^\nu_{\alpha\beta} - \bar{\Gamma}^\nu_{\alpha\beta}\right)\sigma^{\alpha\beta} = 0, \tag{3}$$

where

$$\sigma^{\mu\nu} = \frac{\partial W(\mathfrak{g};\bar{\mathfrak{g}})}{\partial \epsilon_{\mu\nu}} = 2\frac{\partial W(\mathfrak{g};\bar{\mathfrak{g}})}{\partial \mathfrak{g}_{\mu\nu}}, \tag{4}$$

along with the boundary conditions

$$n_\alpha \sigma^{\alpha\beta} = t^\beta, \tag{5}$$

where $t^\beta$ is a boundary traction, and $n_\alpha$ is the unit vector normal to the boundaries. Here $\Gamma$ and $\bar{\Gamma}$ are the Christoffel symbols associated with $\mathfrak{g}$ and $\bar{\mathfrak{g}}$ respectively, and $\bar{\nabla}$ is the covariant derivative with respect to $\bar{\mathfrak{g}}$, namely,

$$\bar{\nabla}_\mu \sigma^{\mu\nu} = \partial_\mu \sigma^{\mu\nu} + \bar{\Gamma}^\mu_{\mu\beta}\sigma^{\beta\nu} + \bar{\Gamma}^\nu_{\mu\beta}\sigma^{\beta\mu}.$$

Equation (3) is a momentum conservation equation, and as such is independent of the specific constitutive law. The constitutive law enters in the relation (4) between the stress and the configuration. The equilibrium equations (3), together with the boundary conditions (5) and the constitutive law (4) form a closed system of equations.

## III. THE INCOMPATIBLE STRESS FUNCTION

The dependent variable whose solution we seek is conventionally taken to be the configuration. In this section we adopt a different approach, and express the elastic problem as a system of equations in which the unknown is the actual metric $\mathfrak{g}$. We focus on two-dimensional problems. It is important to note that one could also consider cases in which the ambient space is non-Euclidean, e.g., the surface of a sphere. In this paper we only consider embeddings in Euclidean plane, in which case the elastic problem is known as the *plane-stress problem*. The case of a non-Euclidean ambient space will be treated in a subsequent publication.

A well-known fact is that any two-dimensional divergence-free tensor field can be expressed as the tensorial action of a curl on the gradient of a scalar function. Here we generalize this property to the generalized Riemannian setting. In Appendix A we show that any stress field solving (3) can be represented as

$$\sigma^{\mu\nu} = \left(\frac{1}{\sqrt{\det\bar{\mathfrak{g}}}}\varepsilon^{\mu\alpha}\right)\left(\frac{1}{\sqrt{\det\mathfrak{g}}}\varepsilon^{\nu\beta}\right)\nabla_\alpha\bar{\nabla}_\beta\psi \tag{6}$$

where $\varepsilon$ is the Levi-Civita anti-symmetric symbol, and $\bar{\nabla}$ and $\nabla$ are the covariant derivative with respect to $\bar{\mathfrak{g}}$ and $\mathfrak{g}$, respectively. We call the scalar function $\psi$ the *Incompatible Stress Function* (ISF). It is a generalization of the Airy stress function for the case of a general Riemannian metric.

Note, however, that (6) involves no approximation, and it solves the full nonlinear equilibrium set of equations (3). A constitutive relation establishes a relation between the actual metric $\mathfrak{g}$ (which determines the strain) and the stress $\sigma$

$$u = F(\sigma), \tag{7}$$

where $F$ specify the constitutive relation. In view of (6), a constitutive relation determines a relation between the ISF and $\mathfrak{g}$

$$\mathfrak{g} = \bar{\mathfrak{g}} + 2F\left(\frac{1}{\sqrt{|\bar{\mathfrak{g}}|\,|\mathfrak{g}|}}\varepsilon^{\mu\alpha}\varepsilon^{\nu\beta}\nabla_\alpha\nabla_\beta\psi\right). \tag{8}$$

Since $\mathfrak{g}$ is an actual metric that corresponds to a planar configuration, it must be Euclidean, and we obtain a geometric constraint on the ISF. We have thus reduced the full elastic problem into that of finding an ISF corresponding to a Euclidean $\mathfrak{g}$. Thus, the elastic problem may be restated as

Find $\psi$ such that $\mathfrak{g}$ given by (8) is Euclidean.

This scheme captures both elastic nonlinearity and geometric incompatibility. In addition, it is valid for any constitutive relation, which only affects the relation between $\psi$ and $\mathfrak{g}$.

### A. HOOKEAN SOLIDS

Consider a Hookean constitutive law (though we could choose other ones),

$$\sigma^{\mu\nu} = \mathcal{A}^{\mu\nu\alpha\beta}u_{\alpha\beta}, \tag{9}$$

where

$$\mathcal{A}^{\alpha\beta\gamma\delta} = \frac{Y}{1+\nu}\left(\frac{\nu}{1-\nu}\bar{\mathfrak{g}}^{\alpha\beta}\bar{\mathfrak{g}}^{\gamma\delta} + \bar{\mathfrak{g}}^{\alpha\gamma}\bar{\mathfrak{g}}^{\beta\delta}\right)$$

is the homogeneous and isotropic elastic tensor, $Y$ is Young's modulus, and $\nu$ is the Poisson ratio. Inverting this expression and using (1), we express the actual metric in terms of the stress,

$$\mathfrak{g}_{\mu\nu} = \bar{\mathfrak{g}}_{\mu\nu} + 2\mathcal{A}_{\mu\nu\alpha\beta}\sigma^{\alpha\beta},$$

where we defined

$$\mathcal{A}_{\alpha\beta\gamma\delta} = \frac{1+\nu}{Y}\left(-\frac{\nu}{1+\nu}\bar{\mathfrak{g}}_{\alpha\beta}\bar{\mathfrak{g}}_{\gamma\delta} + \bar{\mathfrak{g}}_{\alpha\gamma}\bar{\mathfrak{g}}_{\beta\delta}\right).$$

Substitution of (6) results in an expression for the actual metric in terms of the elastic constants, the reference metric, and the ISF,

$$\mathfrak{g}_{\mu\nu} = \bar{\mathfrak{g}}_{\mu\nu} + \frac{2\mathcal{A}_{\mu\nu\alpha\beta}}{\sqrt{\det\bar{\mathfrak{g}}}\,\sqrt{\det\mathfrak{g}}}\varepsilon^{\alpha\gamma}\varepsilon^{\beta\kappa}\nabla_\gamma\bar{\nabla}_\kappa\psi. \tag{10}$$

This expression for $\mathfrak{g}$ is implicit as $\mathfrak{g}$ appears on the right-hand side both in the denominator, and in the covariant derivative $\nabla$ which depends nonlinearly on $\mathfrak{g}$.

### B. Geometric compatibility conditions

In the previous section we obtained an expression for the unknown actual metric $\mathfrak{g}$ in terms of the ISF. This expression embodies both the equilibrium condition (3) and the constitutive law (9). We are then left with the problem of determining the ISF.

Note that not every metric $\mathfrak{g}$ is acceptable. Since the body manifold is embedded in Euclidean space, the actual metric must be Euclidean (the actual metric is by definition the metric $\mathfrak{g}$ that makes the map $f : (\mathcal{B}, \mathfrak{g}) \to (\mathcal{S}, \mathfrak{Eucl.})$ an isometry). In two dimensions, a necessary condition for $\mathfrak{g}$ to be Euclidean is the vanishing of the Gaussian curvature,

$$K_G = 0, \tag{11}$$

which by Gauss' theorem, only depends on the metric and not on the embedding. If the body manifold is simply-connected, then this condition is also sufficient. In many cases, however, one may be interested in other topologies, for example, an annulus. In such cases, a vanishing Gaussian curvature does not guarantee a (globally) Euclidean geometry. In a recent work [20] we showed that an annular manifold can be isometrically embedded in the Euclidean plane if and only if its Gaussian curvature vanishes, and in addition its *monodromy* is trivial. The monodromy is a map from the fundamental group of the manifold to a space of affine transformations,

$$x \to Ax + b$$

where $A$ is a linear transformation and $b$ is a constant vector. The monodromy is trivial if its image contains only the identity, i.e, $A = I$ and $b = 0$.

The condition $A = I$ is equivalent to

$$\oint \kappa_g \, dl = -2\pi, \tag{12}$$

and the condition $b = 0$ is equivalent to

$$\oint \Pi^p_{\gamma(t)} \left(\dot{\gamma}(t)\right) \, dt = 0, \tag{13}$$

where $p$ is an arbitrary reference point, and the integrals are along any closed curve $\gamma$ with winding number 1. Here, $\kappa_g$ is the geodesic curvature along the curve and $\Pi^p_q$ is the parallel transport operator from point $q$ to $p$. The latter is well-defined on locally Euclidean manifolds when $A = I$ (see [20] for details). The physical interpretation of these conditions is that both the Frank and the burgers vectors associated with the intrinsic geometry of the material vanish for every closed curve. The local equation (11), along with the conditions (12) and (13), are compatibility conditions for $\mathfrak{g}$ to be an actual metric of a surface embedded in the Euclidean plane.

Thus, the plane-stress problem can be reformulated as follows: find a metric $\mathfrak{g}$ of the form (10), satisfying the compatibility conditions (11), (12) and (13) and the boundary conditions (5).

## IV. APPROXIMATION METHODS

The plane-stress problem, as reformulated geometrically in the previous section, is still highly nonlinear and not generally solvable by analytical means. For the geometric approach to be of practical interest, approximation methods must be developed. The first step of any systematic perturbative approach is the identification of small parameters.

Since our problem results from a geometric incompatibility, the expansion parameter is expected to quantify the extent of geometric incompatibility. When g is smooth, every open set of sufficiently small diameter can be embedded in Euclidean space "almost isometrically". Physically, this means that a small enough sample has a configuration that is almost strain-free. This suggest that for the case of a smooth reference metric, a natural expansion parameter is a product of the size of the body and a characteristic Riemannian curvature. Generally, the quantification of geometric incompatibility is problem-dependent.

Suppose that $\eta$ is a small dimensionless parameter that measures the amount of geometric incompatibility. Assuming the ISF is analytic in the neighborhood of $\psi = 0$ we expand the it in powers of $\eta$,

$$\psi = \eta\psi^{(1)} + \eta^2\psi^{(2)} + O(\eta^3)$$

Equation (8) induces a similar expansion for $\mathfrak{g}$,

$$\mathfrak{g} = \bar{\mathfrak{g}} + \eta\mathfrak{g}^{(1)} + \eta^2\mathfrak{g}^{(2)} + O(\eta^3)$$

which in turn induces an expansion for the actual Gaussian curvature

$$K_G = \bar{K}_G + \eta K^{(1)} + \eta^2 K^{(2)} + O(\eta^3)$$

For the special case of a Hookean solid, to leading order, we may replace $\mathfrak{g}$ and $\nabla$ on the right hand side of (10) by $\bar{\mathfrak{g}}$ and $\bar{\nabla}$, obtaining

$$\mathfrak{g}_{\mu\nu} = \bar{\mathfrak{g}}_{\mu\nu} + \frac{2\eta}{\det \bar{\mathfrak{g}}} \mathcal{A}_{\mu\nu\alpha\beta}\varepsilon^{\alpha\gamma}\varepsilon^{\beta\kappa}\bar{\nabla}_\gamma\bar{\nabla}_\kappa\psi^{(1)} + O(\eta^2). \tag{14}$$

Having an explicit expression for a first-order approximation for the actual metric, we turn to impose the geometric compatibility conditions. We start with the condition (11) on the curvature. The Gaussian curvature is

$$K_G = \frac{1}{2}\mathfrak{g}^{\alpha\gamma}\mathfrak{g}^{\beta\delta}R_{\alpha\beta\gamma\delta},$$

where $R_{\alpha\beta\gamma\delta}$ is the Riemann curvature tensor,

$$\begin{aligned} R_{\alpha\beta\gamma\delta} = \frac{1}{2}\left(\partial_{\beta\gamma}\mathfrak{g}_{\alpha\delta} + \partial_{\alpha\delta}\mathfrak{g}_{\beta\gamma} - \partial_{\alpha\gamma}\mathfrak{g}_{\beta\delta} - \partial_{\beta\delta}\mathfrak{g}_{\alpha\gamma}\right) + \\ \mathfrak{g}_{\mu\nu}\left(\Gamma^\mu_{\beta\gamma}\Gamma^\nu_{\alpha\delta} - \Gamma^\mu_{\beta\delta}\Gamma^\nu_{\alpha\gamma}\right). \end{aligned}$$

Since expression (14) for $\mathfrak{g}$ is accurate to first-order in $\eta$, we may impose the compatibility condition $K_G = 0$ only to that order. This results in a PDE for the first-order term of the ISF $\psi^{(1)}$,

$$-\frac{1}{Y}\bar{\Delta}\bar{\Delta}\psi^{(1)} - \frac{2\bar{K}_G}{Y}\bar{\Delta}\psi^{(1)} + \bar{K}_G - \frac{1+\nu_p}{Y}\bar{\mathfrak{g}}^{\mu\nu}\left(\partial_\mu\bar{K}_G\right)\left(\partial_\nu\psi^{(1)}\right) = 0. \tag{15}$$

Here $\bar{K}_G$ is the Gaussian curvature associated with the reference metric $\bar{\mathfrak{g}}$, and $\bar{\Delta}$ is the Laplace-Beltrami operator with respect to $\bar{\mathfrak{g}}$,

$$\bar{\Delta} f = \frac{1}{\sqrt{\bar{\mathfrak{g}}}}\partial_\mu\left(\sqrt{\bar{\mathfrak{g}}}\,\bar{\mathfrak{g}}^{\mu\nu}\partial_\nu f\right).$$

Equation (15) together with the boundary conditions and the geometric compatibility conditions completely define the solution $\psi^{(1)}$, up to immaterial gauge transformations.

In classical "compatible" elasticity it is assumed that $\bar{K}_G = 0$. In this case (15) reduces, as expected, to the biharmonic equation, which is the equation satisfied by the classical Airy stress function. Moreover, in **compatible linear** elasticity, the compatibility condition are imposed on the **linearized strain**

$$\frac{\partial^2 \bar{u}_{11}}{\partial x_2^2} - 2\frac{\partial^2 \bar{u}_{12}}{\partial x_1 \partial x_2} + \frac{\partial^2 \bar{u}_{22}}{\partial x_1^2} = 0, \tag{16}$$

which is a linearized approximation to the condition $K_G = 0$ (see Appendix B). In addition to being free of geometric linearization, our approach yields two additional geometric constraints. The condition of trivial monodromy has no immediate analog in the classical approach. The absence of geometric compatibility conditions is noticeable in problems involving a non-trivial topology, where the constants of integration are usually determined by heuristic considerations. For example (see [21]), in disclinations or dislocations, the boundary conditions alone do not determine the solution uniquely. Additional constraints on the displacement field are often imposed arbitrarily. In the current approach, the equations are always fully determined.

### A. Iterative perturbation method

The geometric approach using the ISF allows a perturbative approximation. Once we have solved the equation for $\psi^{(n)}$, we obtain a linear equation for $\psi^{(n+1)}$. In this section we derive the second-order correction.

Going back to (10), the $O(\eta^2)$ equation comprises three terms: (i) A term linear in $\psi^{(2)}$. (ii) The first-order correction for $1/\sqrt{\det \mathfrak{g}}$. (iii) The first-order correction for connection coefficients in the covariant derivatives. The last two terms depend on the leading-order solution, $\psi^{(1)}$.

The detailed calculations are given in Appendix C. We express the actual metric as follows,

$$\mathfrak{g}_{\mu\nu} = \bar{\mathfrak{g}}_{\mu\nu} + \mathfrak{g}^{(1)}_{\mu\nu} + \mathfrak{g}^{(2)}_{\mu\nu} + O(\eta^3), \tag{17}$$

where

$$\mathfrak{g}^{(1)}_{\rho\sigma} = \frac{2}{\det \bar{\mathfrak{g}}} \mathcal{A}_{\rho\sigma\alpha\beta} \varepsilon^{\alpha\mu} \varepsilon^{\beta\nu} \bar{\nabla}_\mu \bar{\nabla}_\nu \psi^{(1)},$$

and

$$\begin{aligned} \mathfrak{g}^{(2)}_{\rho\sigma} = &\frac{2}{\det \bar{\mathfrak{g}}} \mathcal{A}_{\rho\sigma\alpha\beta} \varepsilon^{\alpha\mu} \varepsilon^{\beta\nu} \bar{\nabla}_\mu \bar{\nabla}_\nu \psi^{(2)} \\ &- \frac{1}{\det \bar{\mathfrak{g}}^2} \operatorname{Tr}\left(\bar{\mathfrak{g}} \text{ adj } \mathfrak{g}^{(1)}\right) \mathcal{A}_{\rho\sigma\alpha\beta} \varepsilon^{\alpha\mu} \varepsilon^{\beta\nu} \bar{\nabla}_\mu \bar{\nabla}_\nu \psi^{(1)} \\ &- \frac{2}{\det \bar{\mathfrak{g}}} \mathcal{A}_{\rho\sigma\alpha\beta} \varepsilon^{\alpha\mu} \varepsilon^{\beta\nu} \delta\Gamma^\gamma_{\mu\nu} \partial_\gamma \psi^{(1)}, \end{aligned} \tag{18}$$

with

$$\delta\Gamma^\mu_{\nu\rho} = \left(c^{\mu\beta} \bar{\mathfrak{g}}_{\beta\gamma} \bar{\Gamma}^\gamma_{\nu\rho} + \bar{\mathfrak{g}}^{\mu\beta} \xi_{\beta\nu\rho}\right), \tag{19}$$

$$\xi_{\alpha\beta\gamma} = \frac{1}{2}\left(\partial_\gamma \mathfrak{g}^{(1)}_{\alpha\beta} + \partial_\beta \mathfrak{g}^{(1)}_{\alpha\gamma} - \partial_\alpha \mathfrak{g}^{(1)}_{\beta\gamma}\right),$$

and

$$c^{\mu\nu} = \frac{1}{\det \bar{\mathfrak{g}}}\left(\left(\text{adj } \mathfrak{g}^{(1)}\right)^{\mu\nu} - \operatorname{Tr}\left(\bar{\mathfrak{g}} \text{ adj } \mathfrak{g}^{(1)}\right) \bar{\mathfrak{g}}^{\mu\nu}\right).$$

For a matrix $A$, $\operatorname{adj} A$ denotes its adjugate, which is the transpose of the cofactor matrix.

Having an expression (17) for the actual metric, we write down the geometric compatibility condition (11) up to second order, obtaining an equation for $\psi^{(2)}$. The The equation (11), together with (12) and (13) determine the elastic solution for $\psi^{(2)}$.

## V. EXAMPLES

### A. Disclinations

As a first example we solve the classical problem of a wedge disclination. Classically, a disclination is a defect created by the removal/insertion of a wedge (see Fig. 2). However, a disclination geometry can also be formed by differential growth that induces a volume expansion factor $\varphi(\rho, \theta) = \alpha \log \rho$, where $(\rho, \theta)$ are polar coordinates. The resulting reference metric is [3]

$$\bar{\mathfrak{g}}(\rho, \theta) = \rho^{2\alpha} \begin{pmatrix} 1 & 0 \\ 0 & \rho^2 \end{pmatrix}. \tag{20}$$

By rescaling the radial coordinate,

$$r = \frac{\rho^{1+\alpha}}{1+\alpha}$$

the reference metric takes the more familiar form

$$\bar{\mathfrak{g}}(r, \theta) = \begin{pmatrix} 1 & 0 \\ 0 & q^2 r^2 \end{pmatrix}, \tag{21}$$

where $q = 1 + \alpha$. We assume an annulus of inner radius $r_{\text{in}}$ and outer radius $r_{\text{out}}$, and impose free boundary conditions.

The case $q = 1$ corresponds to a Euclidean annulus. For $q \neq 1$, the reference Gaussian curvature $\bar{K}_G$ is also zero everywhere, i.e., the reference metric is locally-Euclidean. Indeed, a disclination has a cone geometry, with a cone angle

$$\Delta\theta = -2\pi\alpha.$$

The intrinsic geometry of a discliniation can be revealed by allowing a very thin slice of the material to buckle in 3D space. The isometric embedding of (21) is a cone. Therefore the plane-stress state of a material with a disclination-line is equivalent to the stress-state of a flattened cone Fig. 2.

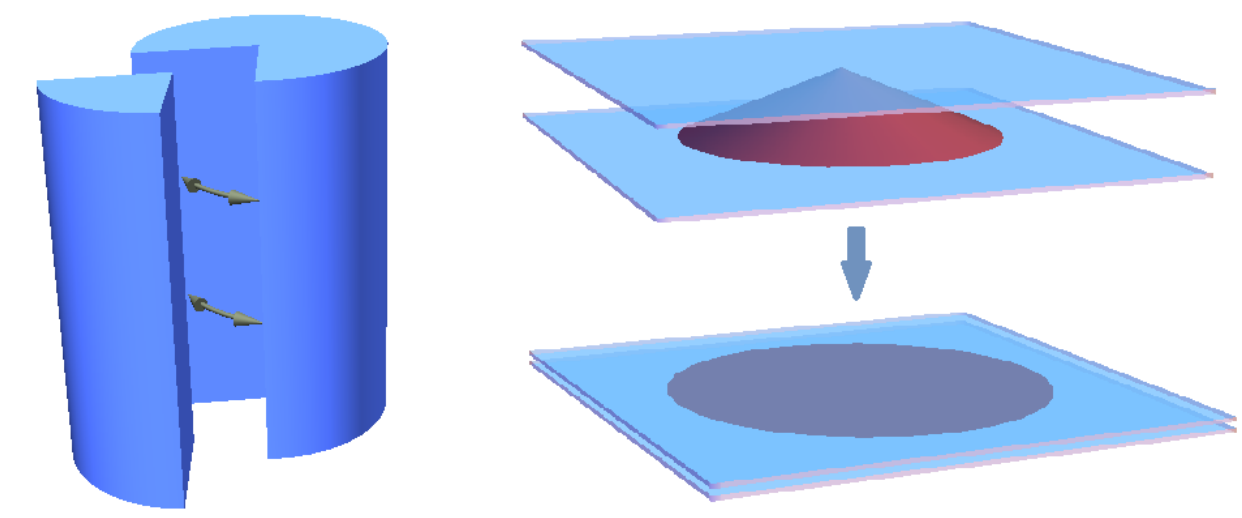

FIG. 2: An illustration of the geometry and the stressed state of a disclination. The stressed state of a 3D material containing a disclination on a cross-section (left panel) is identical to that of a flattened cone (right panel).

### 1. *First-order approximation*

Substituting the reference metric (21) and $\bar{K}_G = 0$ into equation (15) we obtain the biharmonic equation,

$$\bar{\Delta}\bar{\Delta}\psi^{(1)} = 0,$$

where $\bar{\Delta}$ is the Laplace-Beltrami operator. The general solution is the well-known Michell solution [22]. In the case of an axially-symmetric problem, the solution is independent of $\theta$. The general solution is

$$\psi^{(1)}(r,\theta) = A\log r + \frac{Br^2}{2} + \frac{1}{2}Cr^2\left(\log r - \frac{1}{2}\right). \quad (22)$$

where $A$, $B$, and $C$ are constants of integrations.

Classically, at this point one obtain expression for the displacement field and require non zero frank vector. This procedure requires additional linearities and describes *extrinsic disclination*. Instead, next we directly impose geometric compatibility conditions that has no analogue in the classical method.

The geodesic curvature of a circular loop is

$$\kappa_g = \frac{\sqrt{\mathfrak{g}_{rr}\mathfrak{g}_{\theta\theta} - \mathfrak{g}_{r\theta}^2}}{\mathfrak{g}_{\theta\theta}^{3/2}}\Gamma^r_{\theta\theta}.$$

Substituting (22) into (12) we find the linear component of the monodromy to leading order, obtain a first constraint

$$\oint \kappa_g\sqrt{\mathfrak{g}_{\theta\theta}}d\theta = -\frac{4\pi\zeta Cq}{Y} - 2\pi q = -2\pi,$$

hence,

$$C = -\frac{(q-1)Y}{2q}.$$

The vanishing of the translational component of the monodromy, (13), is automatically satisfied by any axially-symmetric solution. To determine the remaining constants $A$ and $B$ we need to impose boundary conditions. By (6), the radial stress component is given by

$$\sigma^{rr} = \frac{1}{q^2r^2}\nabla_\theta\nabla_\theta\psi^{(1)}.$$

To first-order in $\eta$,

$$\sigma^{rr} = -\frac{1}{q^2r^2}\bar{\Gamma}^r_{\theta\theta}\frac{\partial\psi^{(1)}}{\partial r} = \frac{A}{r^2} + B + \frac{(1-q)Y}{2q}\log r.$$

Imposing free boundary conditions,

$$\sigma^{rr}\,|_{r_{\text{in}}} = 0 \qquad \text{and} \qquad \sigma^{rr}\,|_{r_{\text{out}}} = 0$$

we obtain,

$$A = \frac{(1-q)Y}{2q}\cdot\frac{r_{\text{in}}^2 r_{\text{out}}^2\ln(r_{\text{in}}/r_{\text{out}})}{r_{\text{in}}^2 - r_{\text{out}}^2},$$

and

$$B = \frac{(q-1)Y}{2q}\cdot\frac{r_{\text{in}}^2\ln(r_{\text{in}}) - r_{\text{out}}^2\ln(r_{\text{out}})}{r_{\text{in}}^2 - r_{\text{out}}^2}.$$

To conclude, the stress components are

$$\sigma_r^r = \bar{\mathfrak{g}}_{r\mu}\sigma^{\mu r} = \frac{(1-q)Y}{2q}\frac{r_{\text{in}}^2 r_{\text{out}}^2}{r_{\text{in}}^2 - r_{\text{out}}^2}\frac{\ln(r_{\text{in}}/r_{\text{out}})}{r^2} + \frac{(1-q)Y}{2q}\frac{r_{\text{out}}^2\ln(r_{\text{out}}/r) - r_{\text{in}}^2\ln(r_{\text{in}}/r)}{r_{\text{in}}^2 - r_{\text{out}}^2},$$

and

$$\sigma_\theta^\theta = \bar{\mathfrak{g}}_{\theta\mu}\sigma^{\mu\theta} = \frac{(1-q)Y}{2q}\left(1 - \frac{r_{\text{in}}^2 r_{\text{out}}^2}{r_{\text{in}}^2 - r_{\text{out}}^2}\frac{\ln(r_{\text{in}}/r_{\text{out}})}{r^2}\right) + \frac{(1-q)Y}{2q}\left(\frac{r_{\text{out}}^2\ln(r_{\text{out}}/r) - r_{\text{in}}^2\ln(r_{\text{in}}/r)}{r_{\text{in}}^2 - r_{\text{out}}^2}\right).$$

In particular, these expressions identify $(1-q)\cdot(1-r_{\text{in}}/r_{\text{out}})$ as small parameters. That is small incompatibility relates intrinsic geometric properties with the size of the sample.

In Fig. 3 we plot $\sigma_r^r(r)$ for a disclination charge $\Delta\theta = 2\pi/10$ (or $q = 0.9$). Fig. 3(a) compares the exact numerical solution $\sigma_{ex}$ (solid line), the Airy solution $\sigma_{Airy}$ (dashed red line) and our first-order approximation $\sigma_{ISF}^{(1)}$ (dashed green line). To compare accuracies we plot in Fig. 3(b) the normalized deviations from the exact solution, $\overline{\delta\sigma}_{\text{Airy}} \equiv (\sigma_{Airy} - \sigma_{ex})/\max|\sigma_{ex}|$ (dashed red line) and $\overline{\delta\sigma}_{\text{ISF}} \equiv (\sigma_{ISF}^{(1)} - \sigma_{ex})/\max|\sigma_{ex}|$ (dashed green line). Our first-order approximation is more accurate than the Airy solution. A similar picture is observed for the other stress components.

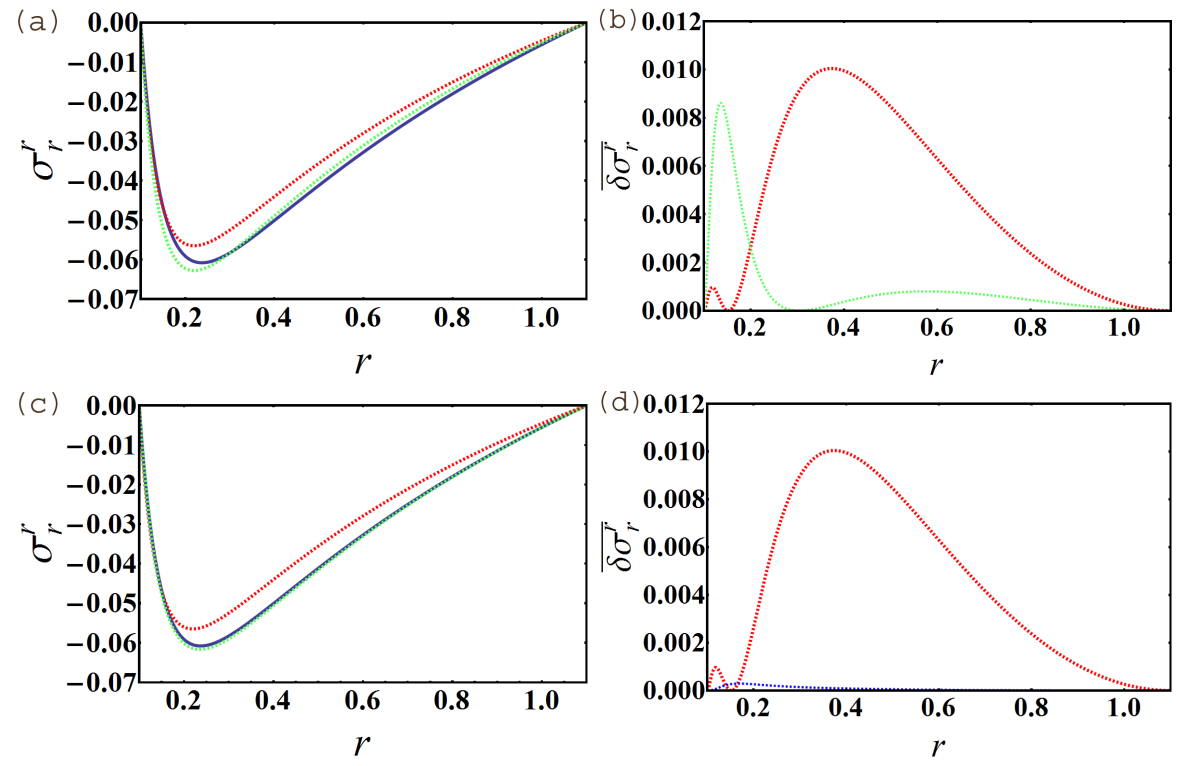

FIG. 3: (a) $\sigma^r_r$ as function of $r$ for a disclination geometry with parameter $q = 0.9$. We compare the exact solution (solid blue line, obtained numerically), the linear Airy solution (dashed red line) and our first-order approximation (green dashed line). (b) Normalized deviations from the exact solution. The red line shows the deviation of the Airy solution and the green line shows the deviation of our first-order approximation. Both are normalized by the maximal value of the exact solution. Figures (c) and (d) are analogous to figures (a) and (b) but use our second-order approximation.

#### 2. *Second-order approximation*

Having calculated $\psi^{(1)}$, we substitute it in (18) and impose the geometric condition (11) up to second order. This results in an equation for $\psi^{(2)}$. Given $\psi^{(1)}$, the Gaussian curvature up to second order is given by

$$K_G = -\frac{1}{4Y}\bar{\Delta}\bar{\Delta}\psi^{(2)}(r) + \frac{(q-1)^2(\nu_p-3)^2}{8q^2r^2} - \frac{(q-1)^2(\nu_p+1)^2 r_{\text{in}}^2 r_{\text{out}}^2 \ln\left(\frac{r_{\text{in}}}{r_{\text{out}}}\right)}{4q^2r^4\left(r_{\text{in}}^2-r_{\text{out}}^2\right)}.$$

The equation $K_G = 0$ is solvable analytically. The constants of integration are determined exactly as in the first-order case.

Fig. 3(c) and (d) are analogous to Fig. 3(a) and (b), except that we replaced the first-order approximation $\sigma^{(1)}$ by the second-order approximation $\sigma^{(2)}$. Within the resolution of the plot, our approximation is almost indistinguishable from the exact solution.

### B. Constant reference Gaussian curvature

In this section we solve the stress-state of a translationally symmetric material whose effective 2D reference metric has a constant Gaussian curvature. The computed stress state is that of a flattened, or pre-buckled, non-Euclidean disc whose reference metric determines a constant (positive or negative) Gaussian curvature $\bar{K}_G$.

It is interesting to note that this solution describes the stress state of a rod that undergoes thermal expansion due to a spatially uniform heat source (see Appendix D). The reference metric is given by:

$$\bar{\mathfrak{g}} = \begin{pmatrix} 1 & 0 \\ 0 & \frac{1}{\bar{K}_G}\sin\left(\sqrt{\bar{K}_G}r\right)^2 \end{pmatrix}, \tag{23}$$

where $(r, \theta)$ are again polar coordinates and $r \in [r_{\text{in}}, r_{\text{out}}]$. Here too, we assume free boundary conditions. Similar to the analogy between disclination and a flattened cone, the stress-state in the current example is equivalent to that of a flattened sphere Fig. 1.

Substituting (23) into (15) we obtain the following equation for $\psi^{(1)}$,

$$-\frac{1}{Y}\bar{\Delta}\bar{\Delta}\psi^{(1)} - \frac{2\bar{K}_G}{Y}\bar{\Delta}\psi^{(1)} + \bar{K}_G = 0.$$

The general axially-symmetric solution is

$$\begin{aligned}\psi^{(1)}(r) = &-\frac{A}{2\bar{K}_G}\cos\left(\sqrt{\bar{K}_G}r\right) + \left(\frac{A}{2\bar{K}_G} - C\right)\tanh^{-1}\left(\cos\left(\sqrt{\bar{K}_G}r\right)\right) \\ &- \frac{B}{2\bar{K}_G}\tanh^{-1}\left(\cos\left(\sqrt{\bar{K}_G}r\right)\right)\cos\left(\sqrt{\bar{K}_G}r\right) \\ &- \frac{Y}{4\bar{K}_G}\ln\left(\sin^2\left(\sqrt{\bar{K}_G}r\right)\right).\end{aligned} \tag{24}$$

The $\sigma^r_r$ component of the stress field is given by

$$\begin{aligned}\sigma^r_r = &\frac{1}{2}B\arctan\left(\cos\left(\sqrt{\bar{K}_G}r\right)\right)\cos\left(\sqrt{\bar{K}_G}r\right) + \bar{K}_G C\frac{\cot\left(\sqrt{\bar{K}_G}r\right)}{\sin\left(\sqrt{\bar{K}_G}r\right)} \\ &+ \frac{1}{2}\left(B - Y - A\cos\left(\sqrt{\bar{K}_G}r\right)\right)\cot\left(\sqrt{\bar{K}}r\right)^2.\end{aligned} \tag{25}$$

As in the previous example, the translational component of the monodromy vanishes for any axially-symmetric solution. The constants of integration are determined by the geometric constraint (12) and the boundary conditions. To assess the accuracy of our solution, we compare it to [5], where the fully nonlinear problem was solved numerically.

In Fig. 4 we present results for $r_{\text{in}} = 0.1$, $r_{\text{out}} = 1.1$ and $\sqrt{\bar{K}_G} = 1/4$. The agreement with the exact solution is within fractions of a percent. As expected, increasing the reference curvature results in a larger error. If needed, second-order accuracy can be achieved using (18).

The same solution as for positive Gaussian reference curvature can be used for the case of a negative reference Gaussian curvature, since it is valid for both negative and positive values of $\bar{K}_G$. In Fig. 5 we compare the first-order approximation for the hyperbolic case with the numerical solution obtained in [5] for various values of $\bar{K}_G$. As in previous examples, the agreement is excellent, and can be improved by taking the second order solution for $\psi^{(2)}$.

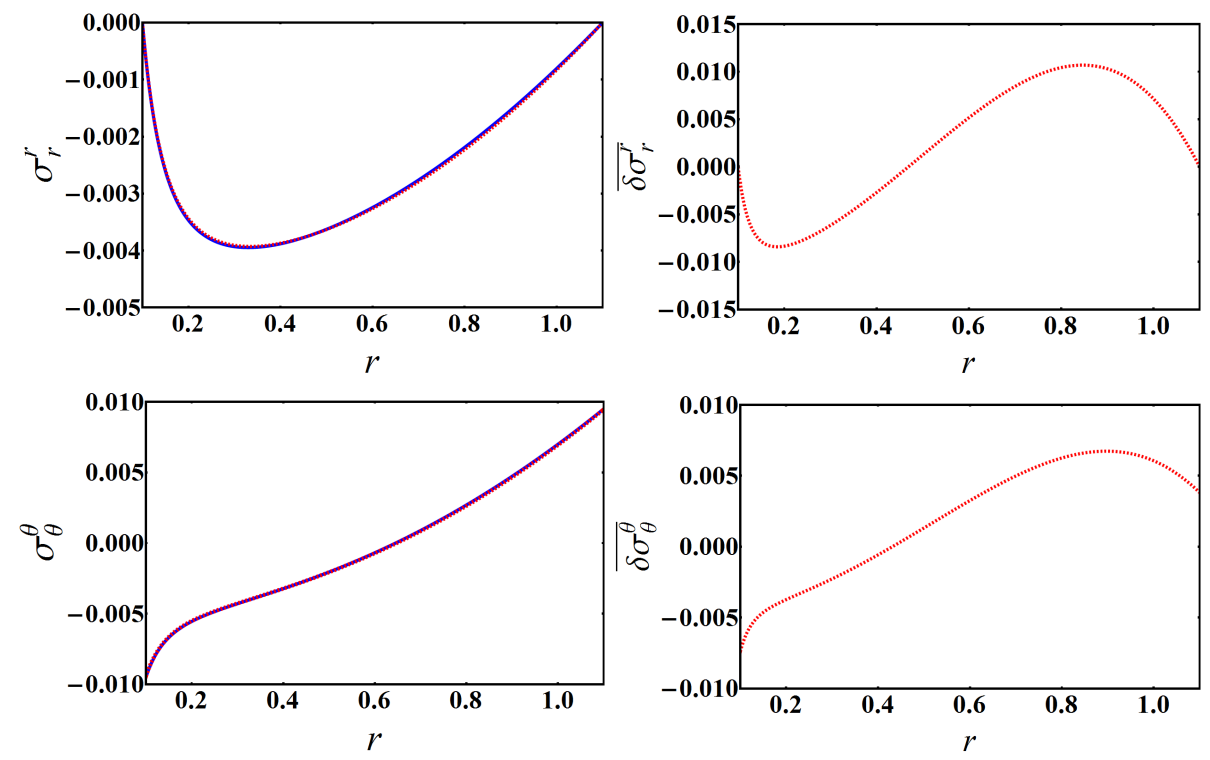

FIG. 4: (a) $\sigma_r^r$ as a function of $r$ for a surface of positive constant reference Gaussian curvature. The parameters are $r_{\text{in}} = 0.1$, $r_{\text{out}} = 1.1$ and $\sqrt{\bar{K}_G} = 1/4$. We compare the exact solution (solid blue line, obtained numerically) to our first-order approximation (red dashed line). (b) Normalized deviations of our first-order approximation from the exact solution. Figures (c) and (d) are analogous to figures (a) and (b) for $\sigma_\theta^\theta$.

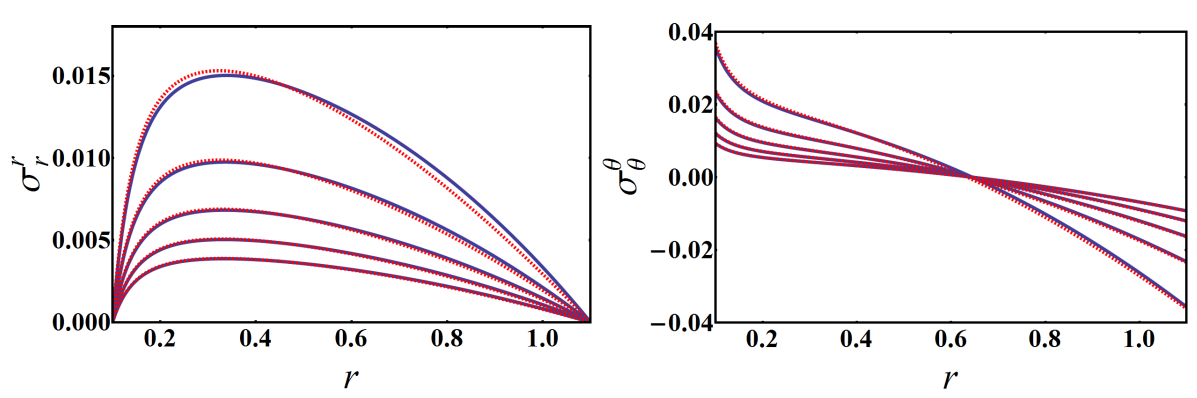

FIG. 5: (a) $\sigma_r^r$ and (b) $\sigma_\theta^\theta$ as functions of $r$ for a surface of constant negative reference Gaussian curvature. Comparison between our first-order approximation (solid line) and the exact solution from [5] (dashed line). The various curves are, top to bottom at the left hand side, for $\bar{K}_G = -1/\Lambda^2$ for $\Lambda = (2, 2.5, 3, 3.5, 4)$.

## VI. CONCLUDING REMARKS

The methods developed in this paper have a wide range of applications, encompassing systems locally characterized by a reference metric. As an example, our approach is relevant to the study of shaping via growth in biological tissue. In this context, the reference metric is prescribed by the underlying biological activity (cell division and expansion). The feedback of mechanical stresses on growth, recently suggested as a growth-regulating mechanism [23], can be included naturally within the formalism by prescribing in addition a (slow) evolution equation for the reference metric.

Further more, as shown in Appendix D, the reference state of a material subjected to temperature gradients can also be described using the reference metric, hence thermoelastic effects can be easily integrated and coupled to the intrinsic geometry of the elastic medium.

Other classes of systems to which this geometric approach can be applied are nematic-elastomers. It was shown that a large set of reference metrics can be prescribed on a nematic elastomer, by designing the appropriate director field (see [24]).

At last, it was recently shown that defects can be defined as singular sources of incompatibility of the reference curvature. This definition, which is appropriate for both ordered and disordered materials, can now be used together with the ISF method to study the mechanics of defects in amorphous materials.

## ACKNOWLEDGMENTS

M.M. and E.S. were supported by the Israel-US Binational Foundation (Grant No. 2008432) and by the European Research Council SoftGrowth project. R. K. was supported by the Israel-US Binational Foundation (Grant No. 2010129) and by the Israel Science Foundation (Grant No. 661/13).